\documentclass[compsoc, conference, a4paper, 10pt, times]{IEEEtran}

\usepackage{cite}
\usepackage{amsmath,amssymb,amsfonts}
\usepackage{algorithmic}
\usepackage{graphicx}
\usepackage{textcomp}
\usepackage{xcolor}
\usepackage{booktabs}
\usepackage[hidelinks]{hyperref}

\usepackage{xspace}

\usepackage{listings}
\usepackage{algorithm}

\newcommand{\ignore}[1]{}
\newcommand{\paragraphb}[1]{\smallskip\noindent{\bf #1.}}
\newcommand{\paragraphi}[1]{\noindent\emph{#1.}}

\newcommand{\combinedtagsz}{16\xspace}

\begin{document}

\title{Memory Tagging: A Memory Efficient Design}

\maketitle

\definecolor{codegreen}{rgb}{0,0.6,0}
\lstdefinestyle{mystyle}{
    commentstyle=\color{codegreen},
    basicstyle=\ttfamily\footnotesize,
    breakatwhitespace=false,         
    breaklines=true,                 
    captionpos=b,                    
    keepspaces=true,
    showspaces=false,                
    showstringspaces=false,
    showtabs=false,                  
    tabsize=2
}

\lstset{style=mystyle}

\begin{abstract}
ARM recently introduced a security feature called {\em Memory Tagging Extension} or MTE,
which is designed to defend against common memory safety vulnerabilities,
such as buffer overflow and use after free. 
In this paper, we examine three aspects of MTE.
First, we survey how modern software systems, such as Glibc, Android, Chrome, Linux, and LLVM, use MTE.
We identify some common weaknesses and propose improvements.
Second, we develop and experiment with an architectural improvement to MTE that 
improves its memory efficiency.
Our design enables longer memory tags, which improves the accuracy of MTE.
Finally, we discuss a number of enhancements to MTE to improve its security 
against certain memory safety attacks. 
\end{abstract}


\section{Introduction}

Memory corruption vulnerabilities continue to be the most widely reported class of vulnerabilities.
A report from the Google Chromium team~\cite{chromeSec} that examined 
912 high or critical severity security bugs since 2015 found that 
about 33\% are due to {\em spatial} bugs, such as buffer overflow, and
about 36\% are due to {\em temporal} bugs, such as use-after-free.
In response, an array of techniques have been deployed to make these vulnerabilities harder to exploit:
non-executable pages~\cite{ms-DEP}, 
address space randomization~\cite{ASLR}, 
stack canaries~\cite{canaries}, 
guard pages,
address sanitization~\cite{asan-atc-12,kasan-desc}, 
control flow integrity~\cite{CFI}, 
and many others. 

In addition to software techniques, 
processor companies have introduced architectural features to defend against memory corruption attacks,
such as a shadow stack and EndBranch in Intel CET~\cite{cet}. 
ARMv8.3 added support for cryptographic control flow integrity~\cite{ccfi}.
This feature, called ARM Pointer Authentication Codes (PAC)~\cite{arm-pac-whitepaper},
embeds a short (say, 16 bits) cryptographic integrity code (MAC) in the top bits of every 64-bit function pointer and return address.  
The MAC is checked before the pointer is dereferenced and an exception is raised if the MAC is invalid.
While this feature ensures that a pointer cannot be overwritten with an arbitrary value~\cite{pac-parts-sec19},
it cannot enforce memory safety. 


ARMv8.5 recently added an important mechanism, called Memory Tagging Extension (MTE)~\cite{arm-mte-whitepaper},
to improve the memory safety of code written in an unsafe language, while requiring no source changes. 
MTE implements a tagging approach to memory access, as proposed in \cite{mem-tagging-google}.
The design is closely related to the design of Hardware-assisted AddressSanitizer (\href{https://clang.llvm.org/docs/HardwareAssistedAddressSanitizerDesign.html}{HWASan}). 

MTE has two parts.  First, all of physical memory is divided into 16-byte granules, and each granule is given an associated 4-bit tag.
These tags are managed by the hardware and are stored in a reserved location in RAM.  
For example, for 64GB of physical memory, the tags take up 2GB of space. 
Second, every pointer and virtual address is modified to contain a 4-bit tag in its top byte. 
This is possible in a 64-bit system because the top byte of a pointer is not used for address translation.
When a pointer is used to access memory, the hardware first verifies that the 4-bit tag embedded in the pointer
matches the 4-bit tag associated with the 16-byte granule being accessed.  
If not, the hardware raises an exception or records that a mismatch occurred.
The reason for using 16-byte granules is explained in~\cite[p.\@ 6]{mem-tagging-google}:
larger granules reduce tag storage, but harm the security of MTE and cause other overheads due to alignment requirements.

\smallskip\noindent
Let us briefly look at a few applications of MTE:
\begin{itemize}[leftmargin=1em,itemsep=1ex]
\item
MTE can reduce the likelihood of a buffer overflow by ensuring that consecutive buffers are tagged with different 4-bit tags.
That way, a pointer into one buffer will trigger an exception if it tries to write to an adjacent buffer.
\item 
MTE can reduce the likelihood of a use-after-free by enhancing the {\tt free} function to change the 4-bit tags associated
with all the granules being freed.  
A dangling pointer will then trigger an exception if it tries to read from or write to the freed space.
\item
MTE can reduce the likelihood of a double-free by making the {\tt free} function check
that the tag in the given pointer matches the tag of the granules being freed, and do nothing if not. 
Since {\tt free} changes the tags of the freed granules, freeing the same space again will do nothing.
\end{itemize}
We stress that these protections are not foolproof and discuss this further
in Section~\ref{sec:usage-survey}.

\smallskip
In Section~\ref{sec:overview} we review the hardware features of MTE 
as well as the Linux kernel changes that were needed to support MTE.
We fill in some details that are missing from ARM's description.

\paragraphb{Our contributions}
Memory allocators need to be modified to take advantage of MTE:
{\tt malloc} needs to tag newly allocated memory and {\tt free} needs to change these tags.
We examined the allocators used in a number of projects and found 
that every project chose to implement a different tagging strategy.
In Section~\ref{sec:usage-survey} we survey the tagging strategies used by 
Glibc, Android Scudo, Linux Slub, Chrome, and LLVM (for tagging stack granules).  

We found significant differences between the tagging strategies in the projects listed above. 
While all of these projects benefit from using MTE, 
some of the strategies used offer much weaker protection than what can be achieved using MTE.
We describe some potential vulnerabilities in all five allocators, 
and describe ways in which the security of these allocators can be improved
with relatively inexpensive modifications. 
We note that currently, there is no support for MTE in iOS, Windows, and the Go runtime.

\smallskip
\paragraphb{Improving MTE}
Next, we look at improving the design of MTE.  
MTE uses 4-bit tags which means that an attacker can guess a required tag with probability $1/16$.
While pointers can accommodate larger tags, the reason 4-bit tags are used is due
to the space required to store tags: 2GB of memory is needed to store tags for a system with 64GB physical memory.
Larger tags, say 16-bit tags, would quadruple the required memory.

In Section~\ref{sec:improvements} we propose a way to support larger tags, 
but without increasing the required memory to store tags.
Our observation is that many allocations are comprised of multiple consecutive granules
{\em where all granules have the same tag}.
This means that we can compress the set of tags using run-length-encoding
and only store the starting point of an allocation, the number of granules it contains, and the assigned tag.
There is no need to store all the tags associated with the allocation in an explicit tag array.
The challenge is that allocations are constantly malloc'ed and free'ed and we need
a data structure that can handle a dynamic run-length-encoding with a fast lookup and a fast update.
We describe such a data structure in Section~\ref{sec:improvements}.
As an additional benefit, this data structure makes it possible to set the tag of many granules at once, 
which can improve the performance of malloc and free.

Next, we implement a modified version of MTE using our data structure.
Our implementation is a software patch to QEMU's emulation of ARMv8.5.
The result is an ARM emulator that implements our modified MTE.

We then experiment with several applications to measure the space savings in tag storage due to our modified MTE.
We experiment with Apache2, md5, FFmpeg, pbzip2, and axel.  
The space savings are significant in applications that mostly make large allocations
and more modest in applications that predominantly make small allocations. 
Figures~\ref{fig:workloads-avg} and~\ref{fig:workload-alloc-cdf} describe the distribution of allocation sizes for each application we experimented with.
Our detailed evaluation in Section~\ref{sec:btree-storage-eval} shows that for all the applications 
we tested, our design supports large tags, even 32-bits long, for 58\% to 99.9\% of RAM (depending on the application) 
with the same storage as MTE's current implementation. 

To summarize, our experiments show that we can support larger tags, 
8, 16, and even 32 bits per granule, for most of RAM, with the same memory overhead for tag storage
as the current implementation of MTE.
The cost is a more complex data structure than the simple tag array used in the current MTE.

Recall that MTE tags are stored in the high-order bits of every pointer,
and this space is also used by PAC (pointer authentication). 
In Section~\ref{sec:additional} we discuss unifying PAC and MTE into a
single value stored in the high-order bits of every pointer.  
We show that this provides protection equivalent to the combination of 16-bit MTE and 16-bit PAC, while only using 16-bits in the pointer.
Finally, we discuss related work in Section~\ref{sec:related}.

\section{Overview of Memory Tagging}
\label{sec:overview}

We begin by describing the detailed operation of MTE
at both the hardware and kernel levels. 
Our description fills in some details that are not described in the
MTE whitepaper~\cite{arm-mte-whitepaper,arm-mte-blog}.

\subsection{Architectural Changes} 

\paragraphb{Memory \& Cache Changes} MTE adds a new memory type called Normal tagged memory. A global tag array stores a 4-bit tag for every 16 bytes (granule) of this memory type: for example, for a machine with 64GB of tagged RAM, the tag array would require 2GB. These tags are stored either in RAM reserved by firmware or in the metadata space available in DRAM chips such as ECC bits. Tags are pre-fetched when cache lines are loaded for tagged memory.  There is also a tag cache used for caching tags. 
Note that tags are applied to physical memory, not to virtual addresses.

\paragraphb{ISA Extensions} MTE adds new instructions for handling tagged memory. 
For every load and store to a granule in tagged memory,
the hardware compares the tag in the top byte of the address register to the tag assigned to the granule being accessed, and can raise an exception in case of mismatch. The {\sf LDG} and {\sf STG} instructions are used to get and set the tag of a granule. The instructions {\sf LDGM} and {\sf STGM} do bulk tag manipulation, but can only be used by system software at Exception Level 1 or higher (e.g. the kernel). These instructions load/store the tags of 1 to 16 consecutive granules at once. 

The {\sf ADDG} and {\sf SUBG} instructions are used for pointer arithmetic: they modify the tag and address separately by an immediate value. 
The {\sf IRG} instruction is used to generate a random tag for a granule. 




\paragraphb{Tag Checking Controls}
MTE defines new system registers that control tag-checking behavior. 

\paragraphi{Global control} The \href{https://developer.arm.com/documentation/ddi0595/2020-12/AArch64-Registers/TCO--Tag-Check-Override}{{\sf TCO}}  register (Tag Check Override) can be used to disable all memory tag checking. Setting the 25th bit of this register to 1 disables tag checking for all loads and stores to tagged memory.

\paragraphi{Handling tag mismatches} When tagging is enabled, there are two modes for handling tag mismatches: synchronous, wherein the hardware throws an exception on tag mismatch; and asynchronous, wherein the CPU accumulates tag mismatch information in a system register ({\sf TFSR\_ELx} or {\sf TFSR\_EL1} for user mode)~\cite{linux-mte-async}. 
In asynchronous mode, the kernel can react to tag faults 
during a context switch or a system call, 
for example, by killing the offending process or thread.
This mode has a lower performance overhead but is less precise than 
synchronous mode. 

\paragraphi{The match-all tag} MTE provides the option of using {\tt 0xF} as a match-all tag, in kernel space. Specifically, if the TCMA bit of the translation control register (\href{https://developer.arm.com/documentation/ddi0595/2020-12/AArch64-Registers/TCR-EL1--Translation-Control-Register--EL1-?lang=en\#fieldset_0-58_58-1}{{\sf TCR\_EL1}}) is enabled, then all memory accesses from EL1 (kernel space), using a pointer tagged with {\tt 0xF}, are not checked for tag equality. This feature is likely inherited from the match-all tag in the software-based kernel address sanitizer (\href{https://www.kernel.org/doc/html/latest/dev-tools/kasan.html}{KASAN}).
It lets the kernel allocate a large buffer, 
without spending the time to set the tags 
for all the granules in the buffer.

\subsection{OS/Kernel Support}

The Linux kernel added support for enabling memory tagging. 
We describe the key updates to the kernel.

\paragraphi{Enabling tagged addresses} 
To enable MTE, a user process must 
call {\rm prctl} with the {\sc pr\_set\_tagged\_addr\_ctrl}
option to be able to pass tagged addresses to the kernel~\cite{linux-prot-mte}.
It must also enable the tagged memory attribute on an address range using a new {\sc prot\_mte} flag for the mmap and mprotect functions. 
These flags must also be set for stack pages when the process starts, to enable tagging for the stack. The tags are initially set to 0 whenever a page is mapped to a process' memory, and preserved on copy on write. 
Clearing tags for a new page is necessary to avoid leaking sensitive information about the previous owner of the page. 

\paragraphi{Configuring tag mismatch behavior} The tag checking mode (synchronous or asynchronous) is selected for each process by calling prctl with the corresponding flag. Tag mismatches are ignored if no mode is specified.  If a thread's checking mode is none or asynchronous, then kernel accesses to user space are not checked for a matching tag. If the mode is synchronous, the kernel makes a best effort to check its accesses to user space~\cite{linux-prot-mte}.

\paragraphi{Forking} When a process forks, the child process inherits the parent's tagging configuration, memory map attributes, and the memory tags.

\paragraphi{Shared Memory} Linux supports using both {\sc map\_shared} and {\sc map\_private} with MTE, but 
tagging is not supported for file-backed memory mappings. With {\sc map\_shared}~\cite{linux-prot-mte}, both memory and its corresponding tags can be shared between processes. Interestingly, if a process shares memory with a forked child using {\sc map\_shared}, 
and the child changes a tag assigned to shared memory, then the parent's pointers to that address will no longer function. 

\paragraphi{Supporting ptrace} Linux has added support for a tracker process to read and write tags of another process using ptrace, with requests 
{\sc ptrace\_peekmtetags} and {\sc ptrace\_pokemtetags} respectively.

\paragraphi{Page Swapping and Hibernation} A new page flag has been added to track whether a page
has been mapped into user space with MTE enabled. 
The tags are saved/restored when the page is swapped out/in respectively. 
Hibernation support uses the same page flag to copy tags out of tag storage and into memory before the hibernation image is created~\cite{linux-mte-page-swap-hibernate}.
Similarly, Linux has added support to save/restore system registers, such as {\sf TFSR\_EL1}, during suspension~\cite{linux-mte-async}.

\subsection{MTE Limitations}
While MTE is a powerful low-overhead approach to memory safety, it does not prevent all memory corruption attacks.
In particular, it does not protect against:

\paragraphi{Intra-object overflows} When an object is allocated, memory for all its sub-fields is tagged with the same value, so that the returned pointer can be used to access the entire object.  Consequently, an intra-object overflow can be used to corrupt data and pointers contained within the object without causing a tag mismatch exception.


\paragraphi{Blind ROP Attacks} One could attack an MTE-based server in a similar fashion to the Blind ROP attack~\cite{brop-hacking-blind}, 
namely determining the tag of a memory granule, by trying multiple guesses and checking if the server crashes and re-starts. 
Since tags are only 4 bits, this will take 16 restarts in expectation. 

\paragraphi{Speculative execution} Attacks based on speculative execution (such as Spectre~\cite{spectre}) could be used to infer the memory tag value of a granule without triggering a tagging violation~\cite{nofat-isca-21}.

These weaknesses have also been pointed out in past works~\cite{tag-guide, msrc-mte-secu-analy, msrc-bluehat-mte-attack}.

\section{How MTE is used in Real-World Systems}
\label{sec:usage-survey}

Next, we look at how software systems have adapted to use memory tagging.
Every project uses a different strategy for implementing tagging,
and we review the security of these variants.



\subsection{Glibc}
\label{sec:mte-impl-glibc}

Glibc, the GNU C Library, updated its implementation of the malloc and free functions to support ARM MTE for tagging heap memory \cite{glibc-mte-code, glibc-mte-allprs}. On every malloc and realloc, Glibc generates a new {\em random non-zero} tag using the {\sf IRG} instruction, 
and assigns it to all the granules in the allocated memory.  
It writes the tag in the top byte of the returned pointer.
On a call to {\tt free}, all the granules belonging to the freed allocation are tagged with~0.
This ensures that pointers to the freed space will no longer work.



Glibc maintains an object's metadata in the 16 bytes preceding the start of the object, and that single granule is always tagged with tag value zero.
This metadata granule, which precedes every heap allocation, 
acts as a guard to prevent an overflow from one allocation to the next.

Lastly, Glibc checks for a double-free, by verifying that the tag in the provided pointer matches the tag of the memory being freed. 
Since the memory tag will be zero after being freed, the pointer's tag will not match on a double free.

\smallskip
\paragraphb{Security}
We point out some weaknesses in this design.
We assume that the attacker has access to the application's source code and binary, 
and the attacker can craft heap pointers with arbitrary tags and addresses.
For simplicity, we assume that the source code of the system being attacked includes pointer arithmetic instructions
that let the attacker manipulate tags in pointers. 
In practice, this is often done using gadgets in the code that do the desired arithmetic.

\smallskip\noindent
Some attacks on Glibc's tagging strategy are possible due to the use of a deterministic tag, namely zero, 
for metadata and for freed space.



\paragraphi{Overwriting the metadata} 
Since the attacker knows that metadata granules are always tagged with zero, 
they could use pointer arithmetic to change the tag of any object pointer to 0, and overwrite the metadata adjacent to the object. 
Listing~\ref{lst:glibc-md-attack-v2} gives a sample source code that is vulnerable to such an attack. The attacker iteratively subtracts one from {\tt arr1} pointer's tag until it becomes zero. Then, the attacker overwrites {\tt arr1}'s metadata at index -16 (16 bytes left from where {\tt arr1} starts). 
A key point is that this attack is guaranteed to work (with probability 1) because the metadata granule is always tagged with zero.  Such exploits can be chained with other vulnerabilities to gain root privilege or allow remote code execution~\cite{glibc-metadata-exploit-pz}.

 \begin{lstlisting}[language=c, caption={glibc Heap Tagging Metadata Vulnerability: The attacker controls the arguments ind, delta, val.}, captionpos=b, label={lst:glibc-md-attack-v2}]
  void src_code(int ind, long delta, char val) {
      char* arr1 = malloc(48);
      // change pointer tag to 0:
      while (((<long>(arr1) >> 56) % 16) > 0)
          arr1 -= delta;
      // overwrite metadata:
      arr1[ind] = val;
  }
  // overwrite arr1's metadata with val:
  src_code(-16, 0x100000000000000, val)
\end{lstlisting}

\ignore{
Benign non-adjacent buffer overflows are probabilistically protected, i.e. they can only succeed with probability $1/16$ (since the memory granule being accessed could have any one of 15 non-zero tag values if allocated or 0 if free/metadata). Similarly, a malicious attacker attempting a non-adjacent overflow can only be successful with probability $1/16$, by trying to guess the memory tag.


 \begin{lstlisting}[language=c, caption=glibc Heap Tagging UAF Vulnerability,captionpos=b, label={lst:glibc-uaf-attack-v2}]
void src_code(long delta, int val) {
    char* arr1 = malloc(48);
    char* arr2 = arr1;
    free(arr1);
    // arr2 is now a dangling pointer
    // set tag to 0:
    while (((reinterpret_cast<long>(arr2) >> 56) % 16) > 0)
        arr2 -= delta;
    // overwrite freed memory:
    arr2[2] = val;
}

src_code(0x100000000000000, val);
\end{lstlisting}
}

\paragraphi{Use after free} 
Since Glibc changes the tags of freed memory to zero, an attacker can again use pointer arithmetic to modify a dangling pointer's tag to 0 and cause a use-after-free vulnerability. As before, the attack is guaranteed to work and can be done using code similar to Listing~\ref{lst:glibc-md-attack-v2}. 
The check for double-free can be similarly bypassed, by modifying the pointer tag to 0, which will then match the tag of freed memory.

Finally, consider a dangling pointer {\tt p} to a freed allocation.  
Once the space is reallocated to another buffer, there is a $1/15$ chance that the new tag will match the old tag (prior to the free).
If that happens then accessing memory using~{\tt p} will not trigger a tag mismatch exception
and enable a use-after-realloc.
The success probability is only $1/15$.

\ignore{
With respect to benign Use-after-realloc's, the library offers probabilistic protection since the tag of the old pointer could match the new memory tag assigned at re-allocation with probability $1/15$ (since a random tag is chosen from 15 possible non-zero tag values). Similarly, an attacker attempting a Use-after-realloc can succeed only with probability $1/15$ by trying to guess the new (non-zero) memory tag.

Lastly, uninitialized memory is zero-tagged and hence benign attempts to access it would be caught deterministically. But, a malicious attacker could access it by modifying a pointer's tag to 0 (similar to Listing~\ref{lst:glibc-uaf-attack-v2}).

We note that we only speculate that these attacks would work but have not actually run them, due to glibc compilation issues with QEMU.
}

\paragraphb{Improvements} 
A natural suggestion is to tag every metadata granule with a {\em random} tag that is different from the tag of the adjacent allocation to the right. 
This prevents the preceding allocation from overflowing into the adjacent allocation
because the overflow pointer cannot match both the tag of the metadata granule and the tag of the adjacent allocation.  
Moreover, it protects the metadata from an overflow from the preceding allocation with probability $15/16 \approx 93\%$.
It also ensures that a simple metadata attack, as in Listing~\ref{lst:glibc-md-attack-v2}, will not succeed with certainty;
it will succeed with probability at most $1/16$ because the attacker must correctly guess the tag of the metadata granule.

Similarly, freed memory could be tagged with a random tag that is different from its current tag value (instead of always tagging freed memory with zero).
As before, it prevents existing pointers from accessing freed memory and prevents a double free. 
More importantly, an attacker can no longer cause a use-after-free with certainty by setting the tag of a dangling pointer to zero.

\subsection{Linux SLUB}
\label{sec:mte-impl-linux}

The Linux kernel uses three different slab-based memory allocators, SLUB, SLOB, and SLAB, with SLUB being the most widely used.
All three include support for software-based memory tagging using \href{https://www.kernel.org/doc/html/latest/dev-tools/kasan.html}{KASAN}.

The SLUB allocator was extended to support MTE, by adding a new hardware tag-based mode in \href{https://www.kernel.org/doc/html/latest/dev-tools/kasan.html}{KASAN} ~\cite{kasan-mte-code, linux-slub-tagging-ppt}.
It uses {\sf IRG} to generate a random tag value for a new memory allocation (via kmalloc), but excludes the tags 0xE and 0xF.
Hence, a new allocation is tagged with one of 14 tags. 
It sets this random tag as the tag for all the newly allocated granules and embeds the tag into the returned pointer. 
Since slab-based allocators round up the allocation size, they can allocate extra granules beyond the requested size. 
These granules are tagged with 0xE.
When memory is freed with a call to {\tt kfree}, all the freed granules are also assigned tag value 0xE. 
The tag 0xE is called the ``no access'' tag, although this ``no access'' property is not enforced by the hardware. 

For internal data structures, the allocator returns a pointer with the match-all tag (0xF) to speed-up memory allocation.
At startup, the kernel sets MTE's  
flag to disable tag checking for pointers with a 0xF tag.
As a result, SLUB need not update the tags of the memory granules allocated for internal data structures, which speeds up memory allocation
but disables tag checking.
Note that no physical memory granule is ever tagged with the match-all tag 0xF.

The allocators implement explicit checks for double free: 
a call to free is rejected unless the pointer tag is either the match-all tag or it matches the current memory tag.


\paragraphb{Security} 
The allocator uses a {\em deterministic} tag (0xE) for freed space, and the same security concerns raised for Glibc with regard to user-after-free (and double-free) apply here as well.

\smallskip
\paragraphi{Exploiting the Match-all tag}
More importantly, there is an opportunity to exploit the match-all tag 0xF of MTE.  
Since Linux enables this match-all feature, any memory location on the kernel's heap can be accessed once a pointer with tag 0xF is crafted. 
An attacker can do so by using pointer arithmetic to modify the tag of any valid pointer to 0xF. 
This again can be done using code similar to Listing~\ref{lst:glibc-md-attack-v2}.
This diminishes the security provided by MTE since a single crafted pointer is enough to obtain read/write access to all kernel memory. 



\subsection{Android Scudo}
\label{sec:mte-impl-scudo}
\href{https://llvm.org/docs/ScudoHardenedAllocator.html}{Scudo} is the default memory allocator used by the Android mobile OS. 
It uses two types of allocators:
a primary allocator for small-sized allocations,  and
a secondary allocator that manages large allocations (at least one page).


\subsubsection{The primary allocator} This is a size-class-based allocator, meaning that it has blocks of memory reserved for objects of specific sizes (also called size classes). Each block can have multiple chunks, with each chunk serving one allocation. It aligns allocations and metadata to 16 bytes, to support tagging granules. At malloc, a new random non-zero tag is generated, and all the allocated granules and the returned pointer are tagged with that value. Similarly, at free, a random non-zero tag is generated, and all the freed granules are tagged with it~\cite{scudo-heap-llvm-tagging-desc, scudo-mte-primary-code}. 
Note that this avoids deterministic tags for chunks. 

Interestingly the allocator implements the following optimization: when malloc re-allocates memory that was previously freed, and the new allocation has the same start address as the previous allocation, malloc re-uses the random tag (that was assigned at free) and assigns it to the pointer. 
If the new allocation is larger than the old one, it tags the additional granules with the same tag.
If the new allocation is smaller, it tags the granule at the very end of the new allocation with zero. 
This avoids the overhead of generating another tag and re-tagging all the granules.

Scudo places 16 bytes of metadata (chunk header) in front of every allocation. This granule is tagged with zero,
but Scudo uses a 16-bit checksum to protect it, stored in the last two bytes of the header. 
The checksum is verified every time the header is accessed, to detect potential corruption.
This metadata granule acts as a guard to prevent a buffer overflow from the preceding allocation.


Scudo provides additional features to enhance security. 
First, it tags the granule just following every allocation with zero. 
This acts as a guard to catch a buffer overflow emanating from the allocation.
The second feature is optional and controls how chunks are tagged. 
If turned on, Scudo generates odd tags for allocations in odd-numbered chunks and even tags for allocations in even-numbered chunks. 

\paragraphb{Security} 
The odd-even tagging feature protects against benign overflows on either side, since the allocations on either side would have a different parity tag. But, a malicious attacker could guess the tags of adjacent allocations with probability $1/8$, since the parity of an object gives away the parity of its adjacent objects in the same size class.


\paragraphi{Temporal Safety} Since Scudo re-tags freed granules with a new random tag at free, any use-after-free attempt (benign or malicious) will be caught with probability $15/16 \approx 93\%$ if random tagging is used, and with $7/8 \approx 87\%$ probability if odd-even tagging is used. 
%
More generally, the odd-even feature provides a security tradeoff: 
it ensures protection for a benign overflow into an adjacent allocation, 
but worsens the probability of catching a use-after-free and a benign use-after-realloc from $93\%$ to $87\%$.

Listing \ref{lst:scudo-uaf-attack} gives an example source code that is vulnerable to a use-after-free attack, 
caused by the optimization where Scudo re-uses the tag generated at free. 
The code first makes a big allocation {\tt a1}, which is then freed. 
Another allocation {\tt a2}, smaller than {\tt a1}, is made immediately, and if {\tt a2} is assigned to {\tt a1}'s freed memory, 
then the {\tt a2} pointer tag will match the tag of all the 40 freed granules. 
The allocator will re-tag the granule right next to {\tt a2} (the 31st granule) with zero, 
but an attacker can use {\tt a2} to overflow into granules 32 to 40 since they are not retagged.

\noindent
\begin{minipage}[t]{.45\textwidth}
\begin{lstlisting}[language=c, caption={Scudo Use-after-free Vulnerability. The attack works for all values of ind between 496 and 639 (32nd to 40th granule).}, captionpos=b, label={lst:scudo-uaf-attack}]
  void src_code(int ind, char val)
  {
      char* a1 = malloc(640);
      free(a1);
      char* a2 = malloc(480);   // a2 == a1
      // overflow into a1's freed granules 
      arr2[ind] = val;
  }
  // attacker calls:
  src_code(500, val);
\end{lstlisting}
\end{minipage}


\paragraphb{Potential Improvements} The use-after-free attack is possible because re-allocating a subset of freed memory does not re-tag the affected granules. To avoid this, it is important to generate a new random tag at malloc and use it to either tag the re-allocated granules or to tag the remaining free granules that remain free.

\subsubsection{Secondary Allocator} Since this allocator is only used for large allocations, Scudo does not tag allocated memory due to the large overhead of setting the tags for all the granules during malloc and free.
Instead, the allocator uses the following protections: it protects the allocation with a non-readable (using {\sc prot\_none}) guard page on the right.
The memory between the beginning of the page and the start of the allocation is tagged with a random non-zero tag.
This acts as a guard to prevent an overflow to the left, since all allocations are 0-tagged.
To protect against temporal attacks, it sets freed pages as non-readable (using {\sc prot\_none}) to catch attempts to access freed memory~\cite{scudo-secAllo-code}.

\ignore{
\paragraphb{Security} 
\paragraphi{Spatial Safety} Since there are guard pages on the right of every allocation marked with {\sc prot\_none} (i.e. inaccessible memory), adjacent overflows will be caught deterministically. In terms of overflowing towards the left, a benign attempt will be caught deterministically since the memory on the left of every allocation is tagged with a non-zero tag, which will not match with the allocation's 0-tagged pointer. But a malicious attempt to overflow to the left could succeed with a 1 in 15 chance (if the attacker correctly guesses the non-zero tag). Non - adjacent overflows into other allocations will succeed since all allocations are tagged with 0.

\paragraphi{Temporal Safety} Since freed memory is marked with {\sc prot\_none} (i.e. inaccessible memory), a use-after-free attempt, whether benign or malicious, will be caught deterministically. 

If memory is reallocated, then, a (0-tagged) dangling pointer would still be able to access the new object since it would also be tagged with zero.

}
\subsection{Chrome PartitionAlloc}
\label{sec:mte-impl-chrome}
PartitionAlloc is the heap memory manager used by both Chrome and Chromium browsers.
This allocator splits memory into different partitions, each reserved for objects of a specific type or size.
Different partitions exist in different regions of the process address space, and there is a guard page on both sides of each partition. 

The allocator uses MTE tagging as follows: at every malloc call (i.e. new heap variable), it generates a random tag, which it assigns to all the allocated memory granules, and to the returned pointer. At free, it increments the tag by 1 for all the freed granules~\cite{chromium-ptalloc-mte}. When re-allocating previously freed memory, it re-uses the incremented tag, and simply assigns it to the returned pointer. 
This optimization improves malloc time. Larger allocations have guard pages on both sides.


The chrome team implemented an interesting mechanism~\cite{chrome-security-blog-uaf-scan} on top of MTE to further defend against use-after-free. 
Their concern is that, due to the ``increment by one'' tagging strategy, 
an attacker can wait for 16 malloc/free cycles of a memory location until the tag for that memory cycles back to an old tag value.
A dangling pointer can be used at that time without triggering a tag mismatch exception. 
To prevent this they use quarantining: when the 4-bit tag of a memory location reaches zero, that location is placed in a quarantine list and is not re-allocated until a periodic heap scan checks that no dangling pointers are pointing to this location.
If so, the location is removed from the quarantine list and can be re-allocated.

\paragraphb{Security} 
Due to the deterministic behavior at free (increment by one), 
an attacker can cause a use-after-free, with probability 1, by incrementing the tag of a dangling pointer by one.
Listing \ref{lst:chrome-uaf-attack} gives an example vulnerable code. 
The attacker uses pointer arithmetic to increment the tag of the dangling pointer {\tt arr2} by 1. Then, the pointer can be used to access freed memory.
Since the incremented tag is also re-used when re-allocating freed memory, an attacker could also cause a Use-after-realloc with an old dangling pointer, similar to Listing \ref{lst:chrome-uaf-attack}. 

\begin{lstlisting}[language=c, caption={Chromium use-after-free due to increment-by-1 tagging strategy.},captionpos=b, label={lst:chrome-uaf-attack}]
  void src_code(long delta, char val) {
      char* arr1 = malloc(64);
      char* arr2 = arr1;
      free(arr1);
      // arr2 is now a dangling pointer
      // increment the tag:
      arr2 += delta;
      // access freed memory:
      arr2[0] = val;
  }
  src_code(0x100000000000000, val);
\end{lstlisting}



\ignore{
Listing \ref{lst:chrome-uar-attack} shows example code on which this attack would work - we assume that {\tt arr3} re-uses the memory that was originally allocated to {\tt arr1}. The attacker can increment the dangling pointer {\tt arr2}'s tag, and can use it to access the new object. 
}


\paragraphb{An Improvement} 
The increment-by-one strategy ensures that all 16 tags are used before the tag recycles to an old value.
A better way to do this is to choose a {\em random odd} 4-bit offset $\delta$ at startup and 
then have the {\tt free} function increment the tag by $\delta$ (mod 16).
This ensures that all 16 tags are used before the tag recycles, as before, but because the attacker does not know $\delta$,
it is harder to mount a use-after-free that succeeds with probability one. To further improve memory safety, the allocator could maintain a small array of {\em random odd} $\delta$s and the low-order bits of the starting address of the allocation could be used to determine which delta gets used when free-ing the allocation.

An alternative approach is to assign {\em random} tags at free, different from the current tag value. 
This will provide a similar security guarantee as the previous paragraph, but with a lower memory overhead.
\subsection{LLVM Stack Tagging}
\label{sec:mte-impl-llvm}
LLVM, the backend used with clang, has been updated to assign MTE tags to stack variables in the function prologue~\cite{llvm-mte-stack-desc, llvm-mte-stack-code, scudo-heap-llvm-tagging-desc}. Similar to the Chrome allocator discussed in the previous section, LLVM uses an increment-by-one strategy to tag stack variables, which enables an overflow attack on stack variables that succeeds with probability one. We were able to successfully run this attack in an MTE-enabled VM in QEMU.
We discuss the implementation, the attack, and our suggested improvements in Appendix~\ref{sec:app-llvm}.

\section{Improving MTE's memory efficiency}
\label{sec:improvements}  



\label{sec:improve-btree}

Since MTE uses 4-bit tags, there is a 1 in 16 chance that a random rogue pointer will contain the correct tag. 
While there is room in the high-order bits of a 64-bit pointer to store longer tags~\cite{linux-tag-addr-abi,arm-pac-whitepaper},
the reason MTE uses such short tags is due to the overhead of storing all tags in memory.
With a 4-bit tag for every 16-byte granule, this overhead is already 3.25\%. 
Concretely, the tag array takes 2GB for a machine with 64GB RAM. 

We propose to improve this significantly by compressing the tag storage using a succinct data structure instead of a tag array.
This lets us use longer tags without taking up additional storage in RAM. 



\begin{figure}[t]
\centering
 \includegraphics[width=0.48\textwidth]{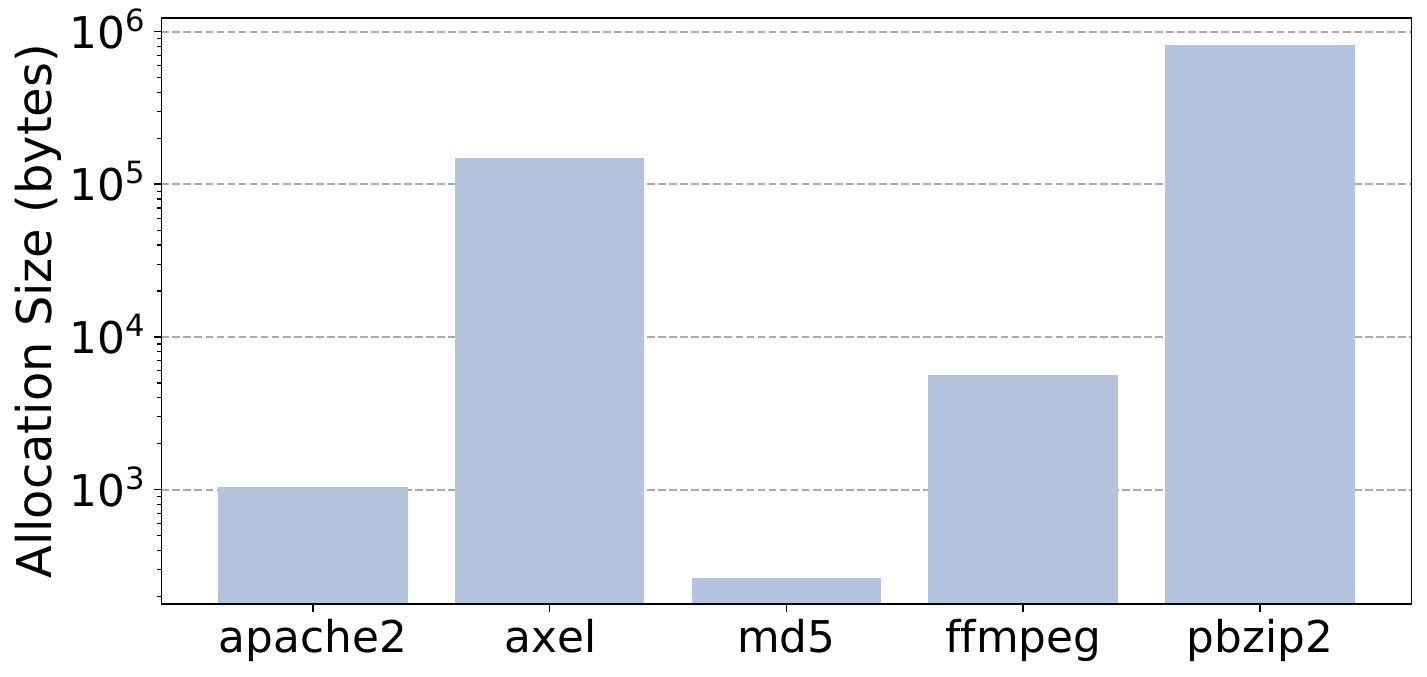}
 \vspace{-10pt}
    \caption{Average allocation size of real-world workloads. The y-axis is in bytes, in log scale.}
\label{fig:workloads-avg}
\vspace{-10pt}
\end{figure}

\subsection{Our approach}
\label{sec:btree-design}

\begin{figure}[t]
\centering
 \includegraphics[width=0.47\textwidth]{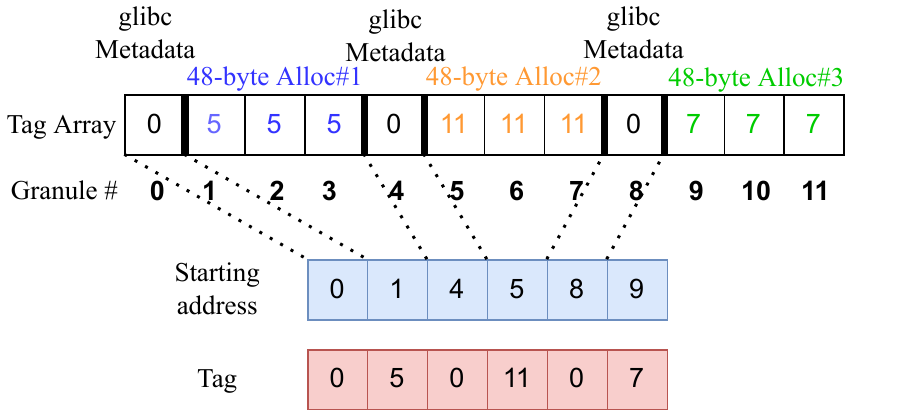}
    \caption{Run Length Encoding Example. We have 3 allocations, each of size 48 bytes (three granules), with one granule of Glibc metadata before every allocation.}
\label{fig:rle-example}
\end{figure}

For a large allocation, say 160 bytes, it can be wasteful to store the same tag value in the tag array for all 10 granules in this allocation. 
Figure~\ref{fig:workloads-avg} shows the average allocation sizes for five real-world workloads running on an Arm64v8 VM. 
We observe that the average size lies between 265 and 818859 bytes (17 to 51179 granules).
This means that a lot of space is wasted by storing the same tag 17 to 51179 times on average.
We could instead use Run Length Encoding (RLE) to store the starting address and tag value for each run, 
as shown in Figure \ref{fig:rle-example}. 
Each continuous set of granules with the same tag is called a run: the example in Figure~\ref{fig:rle-example} has 6 runs.

Storing the RLE encoding instead of an explicit tag array greatly reduces the memory space needed to store tags. 
Consequently, we could either store 4-bit tags in less memory, or use larger tags, say 8, 16, or even 32 bits per granule, 
without expanding the total memory used by MTE. 
However, an RLE-based data structure is useful only if 
(i) reading the tag of a granule is fast (needed in every load and store instruction), and
(ii) maintaining the data structure as tags are updated is efficient (needed in every memory allocation/de-allocation).

\begin{figure}[t]
\centering
 \includegraphics[width=0.25\textwidth]{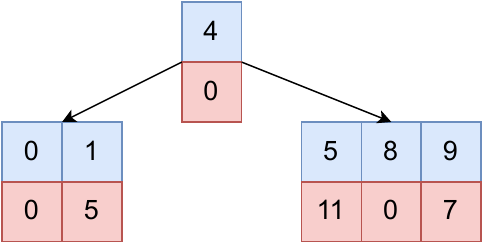}
    \caption{BTree corresponding to the tag array in Figure \ref{fig:rle-example}}
\label{fig:rle-btree-example}
\end{figure}

Our approach supports both properties, while maintaining low memory overhead. 
We do so by using a suitable BTree to store the starting addresses of all runs, along with their corresponding tag values. 
The starting addresses are used as the keys for the tree. 
Figure \ref{fig:rle-btree-example} shows a 4,5-BTree for the tag array in Figure~\ref{fig:rle-example}. 
With this data structure, accessing and modifying the tag of a random granule takes time proportional to the depth of the BTree.
As the number of runs stored in the tree increases, so does the depth of the tree.
As we will see in \S\ref{sec:btree-storage-eval}, the depth of the BTree stays quite small in most real-world workloads. 
Hence, we are able to support larger tags without exceeding the memory overhead of a 4-bit tag array, 
while supporting fast tag access and update. 

With this approach, every pointer will need to store a larger tag in its high-order bits.
We explain how to do that in Section~\ref{sec:btree-discussion}.

In Section~\ref{sec:usage-survey} we saw that allocators try to improve the running time of malloc and free by
minimizing the number of times that they update the tags associated with the granules of a large buffer.  
The BTree data structure has the added benefit that a single update to the tree
can update the tags of many consecutive granules.
This may make it unnecessary to employ the security-harming optimizations used by some allocators.

\paragraphb{Using a per-page BTree}
There are three difficulties with using a global BTree for the whole of RAM. 
\begin{itemize}[leftmargin=1em]
\item Concurrency:
If two processes own adjacent physical memory pages, then it becomes difficult to concurrently update tags for their granules.
This is because the runs for these processes might belong to the same sub-tree or the same node, and hence we might need to use extensive locking. 
While lock-free implementations of BTrees exist~\cite{bw-tree-msr}, they are inefficient and complex to implement~\cite{bw-tree-hard-sigmod}. 

\item Page swaps: 
Whenever the OS swaps a page between memory and disk, the tags corresponding to that page also need to be swapped. 
With a global BTree, this will require reading the tag of every granule in the page, writing it to disk, 
and then updating the global BTree with the tags of all the granules read from disk. 
This can be an expensive operation and combined with the first issue, can affect adjacent pages as well. 

\item An attack:
A global BTree is susceptible to an algorithmic complexity attack~\cite{algo-comp-attacks}, 
where a single process can make a large number of small allocations to increase the size and depth of the tree, 
thereby slowing down all the other processes on the machine. 
For example, if process~1 and process~2 both have one page each, and process~1 only has one allocation, 
then the depth of the BTree would be 1 if process~2 only made one allocation.
However, the depth of the global tree increases to five if process~2 were to make 256 16-byte allocations on its page.
\end{itemize}

\medskip\noindent
To address the difficulties with a global BTree, we opt for a separate BTree for every page. 
This allows different processes to concurrently access and modify the BTrees corresponding to their pages (except for when pages are shared among processes).
Page swapping can be easily done by copying the corresponding BTrees in or out of memory. 
Lastly, a process with a large number of allocations per page will only slow down other processes sharing the same page. 
The only downside is that a per-page data structure takes up a bit more memory than a global BTree, because allocations larger than a page are now split into multiple runs, i.e. one run in the BTree of each page.
But, our experiments show that we can still support large tags without expanding the memory used by MTE. 

\paragraphb{An adaptive data structure} 
The BTree structure can greatly reduce the space requirements when there are few allocations on a page, 
namely when the average allocation size is large. 
However, the BTree may grow too large if a process makes a large number of small allocations. 
If left unchecked, this can increase the memory overhead and slow down accesses and updates to memory tags for that page,
as explained in Section~\ref{sec:btree-analysis}.

To prevent this from happening, we take an adaptive approach.
If the memory usage of the BTree for a page crosses a certain threshold, 
the system switches that page to an explicit tag array that stores 4-bit tags, as in the current MTE implementation. 
This bounds the memory overhead of our design.

\subsection{Implementation}
\label{sec:btree-impl}

We now describe the implementation of our per-page adaptive BTree data structure. 
We aim to support larger tags with the same memory overhead as an explicit 4-bit tag array. 
Our implementation supports 4-bit, 8-bit, 16-bit, and 32-bit tags.
The desired tag size can be chosen at boot time.

An alternative design could instead continue to use 4-bit tags, as currently used by MTE,
and focus on reducing the overall memory requirements.
We do not do that since this would make it much harder to locate the metadata for each page in memory.
Instead, we place the metadata for each page in a fixed location in RAM and aim to use larger tags. 

\medskip
Recall that the default page size for Linux on AArch64 is 4KB.
We assume all pages are of this size, and discuss the impact of variable page sizes in Section~\ref{sec:btree-discussion}.

\smallskip
\paragraphb{Tag Storage}
We reserve a buffer of the same size as the 4-bit tag array for the whole RAM, i.e. 2GB for a machine with 64GB RAM. We fit our BTree data structure with large tags, within this buffer. The BTree for each 4KB page gets 128 bytes (equal to the tag array space for the page), and the tree for the $i$th page is contained between [128i, 128(i+1)] bytes of the buffer.

 \begin{lstlisting}[language=c, caption={The struct defining a node in our per-page BTree for 4KB pages.}, label={lst:btree-node}]
  struct BTreeNode {
    unsigned char keys[4];
    unsigned char children[5];
    unsigned char parent;
    unsigned char tags[TAGARR_NUM_BYTES];
    char count;
  };
\end{lstlisting}

\paragraphb{BTree Implementation}
We build a 4,5-BTree for each page, consisting of nodes as defined by the struct in Listing~\ref{lst:btree-node}. Each node contains the following fields: 
(i) an array of size 4 for storing upto 4 keys. 
Since each key is the start address (granule number) of a run, and there are 256 granules in a 4KB page, 
the key is in the range 0 to 255, and can fit in an 8-bit char. 
(ii) an array of child pointers, of size five, since each node can have up to five children. Each child pointer points to another node in the BTree, and we store the pointer as a relative offset from the starting address of the buffer for the BTree (which is $128i$ for the $i$th page). 
So, the child pointer is in the range 0 to 127, and we store it in an 8-bit char.
(iii) a parent pointer, which also stores the relative offset instead of the absolute address, and hence fits in an 8-bit char. 
(iv) the tags corresponding to each key. The value {\tt TAGARR\_NUM\_BYTES} depends on the tag size: 
we need 2 bytes to store four 4-bit tags, and 4, 8, and 16 bytes to store four 8-bit, 16-bit, and 32-bit tags respectively. 
Lastly, (v) we store the number of keys in the node in a separate field called count. 

Each node in the per-page BTree takes a total of 13, 15, 19, and 27 bytes if 4-bit, 8-bit, 16-bit, and 32-bit tags are used respectively. 
This means that our per-page BTree can only have between 4 and 9 nodes for it to fit within the 128 bytes of reserved metadata space for the page.

For each BTree, we also store additional metadata to manage the space used by the nodes within the buffer (detailed in Appendix~\ref{sec:app-impl}).

\ignore{  
\paragraphi{Other Metadata}
For each BTree, we also store additional metadata: the root pointer, as a 1-byte value, denoting the relative offset of the root node from the start of the buffer for this BTree. To manage the allocation and de-allocation of nodes as the tree grows, we use a 2-byte bit array to denote which nodes in the 128-byte buffer are free. 16 bits are enough since only 4-9 nodes can fit within the 128 bytes buffer. Lastly, we have an additional 1-bit flag for each page, denoting whether we are using a BTree or an array for storing tags for that page.

\paragraphb{Initialization}
We initialize the BTree data structure with the root at offset 0, and all the bits in the free nodes bit-array (except 0th) are marked free. 
The root only has one key: 0, with tag value 0, meaning that the whole page is tagged with 0.
}

\paragraphb{Handling Tag Accesses}
We implement the {\sf STG} and {\sf LDG} functions, i.e. setting and getting the tag of a single granule, by using BTree algorithms for inserting/deleting keys while keeping the tree balanced. The {\sf STG} function additionally tries to merge adjacent runs with the same tag, to reduce the memory overhead.
The {\sf LDG} function simply finds the largest key left to the granule (this is the starting address of the run that the granule belongs to), and returns its tag. 

\paragraphb{Switching to a 4-bit Tag Array}
The system monitors the total space used by the BTree and switches to the 4-bit tag array if the space usage exceeds the max buffer size of 128 bytes. 
If that happens, the system switches a 1-bit flag denoting that the page's tags are now stored in a tag array. 
It iterates through all the allocations stored in the BTree and stores the lower 4 bits of each tag in the explicit tag array. 
Hence, the total space used for storing all the tags for the page never exceeds 128 bytes, as in the current MTE.

We stress that every pointer continues to store the full (4, 8, 16, or 32-bit) tag in its high-order bits. 
The MMU decides how many high-order bits to use based on the
current state of the page being accessed: a BTree or a tag array.
If the page uses a 4-bit tag array, then only the lower four bits of the pointer tag will be used when checking the tag.
We note that it is possible that two objects which had different tags might end up having the same tag after the switch if the lower 4 bits of their tags were equal. This still provides security equivalent to using 4-bit tags.

Once the system shifts a page from a BTree to a \mbox{4-bit} tag array, 
it cannot switch back to the BTree that holds larger tags.
This is because the data required to move back is no longer available in memory. 
The system will reset the page back to a BTree only once the page is released to the operating system. 

\subsection{Space usage analysis}
\label{sec:btree-analysis}
Next, we analyze the space usage and performance of the adaptive BTree design.
We focus on 4KB pages.

Let $g$ denote the total number of granules on the page. Since each page is 4KB, we get $g = 256$. Let $n$ be the number of runs on the page, wherein each run refers to a contiguous set of granules all tagged with the same tag value. For example, each allocation tagged randomly will form a run, and for Glibc, each metadata granule with tag 0 will form a separate run between allocations. Let $t$ denote the tag length (in bytes) for tags being stored in the BTree. This can be 0.5, 1, 2, or 4 bytes. Let $d$ denote the depth of the BTree storing the starting addresses and tags for all $n$ runs. Let $m$ be the number of nodes and $S^{BTree}_t$ be the total space used to store the tree for $t$-byte tags. 

\paragraphb{Space Usage} 
The total space used by each BTree node is $11 + 4t$ bytes, e.g. 13 bytes for 4-bit tags.
Accounting for all $m$ nodes, the metadata, and the fact that each node has at least 2 runs and at most 4 runs, we get the following inequality for the total space used by the BTree (more details in Appendix~\ref{sec:app-analysis}) :
\begin{equation}
\label{eq:btree-space-bounds} 
  \frac{n}{4}(11 + 4t) + 3.125 \leq S^{BTree}_t \leq \frac{n+1}{2}(11 + 4t) + 3.125
\end{equation}
For example, for a 4KB page with 4-bit tags ($t = 0.5$), we consider two cases:
one with $n=2$ runs on the page, and one with $n=256$ runs on the page. 
\begin{itemize}[leftmargin=1em]
\item 
For $n=2$ runs: the tree has just one node, and the total BTree space is 16.125 bytes.
This is $7.9\times$ smaller than the 4-bit tag array space for the page (128 bytes). 

\item
For $n=256$ runs (the worst case number of runs): using Eq.~\eqref{eq:btree-space-bounds} we have $S_t^{BTree} \geq 835$ bytes, 
meaning that the tree will not fit within the allotted 128 bytes.
\end{itemize}

\paragraphb{Switching to Tag Array} 
For each page, the system switches from a BTree to a tag array if the space usage of the BTree exceeds 128 bytes. 
Considering that each node takes $11 + 4t$ bytes, only $m \leq 9, 8, 6$ and $4$ nodes can fit within 128 bytes, 
for 4-bit, 8-bit, 16-bit, and, 32-bit tags respectively. 
Since each node can only store a maximum of 4 runs, the maximum number of runs that can fit within 128 bytes is 
\begin{equation} \label{eq:th}
  n \leq 36, \quad  n \leq 32, \quad n \leq 24, \quad n \leq 16, 
\end{equation}
for 4-bit, 8-bit, 16-bit, and, 32-bit tags respectively. 
The system will switch the page to a tag array if the number of runs $n$ on the page exceeds these thresholds at any point during program execution.

The switch thresholds in \eqref{eq:th} decrease as tag size increases.
Hence, one might expect that larger tags will cause more pages to switch to a tag array. 
We test this out by experiment.







 


\paragraphb{Performance} 
The running time of the {\sf STG} and {\sf LDG} instructions depends on the depth of the tree. 
Given the size constraints, the BTree depth is at most $3$ for 4-bit tags,
and at most $2$ for larger tags.  
Hence, tree reads and updates are quite fast.
Note that we are unable to experiment with a true ARM hardware implementation of the system.

\subsection{Evaluation and Experiments}
\label{sec:btree-storage-eval}

To compare the real-world space usage of our adaptive BTree to a tag array, we collected the complete memory allocation trace of five commonly used programs. 
We ran all programs on a t4g.small AWS VM, which has 2 arm64v8 CPUs and 2 GB RAM. 
We used Valgrind to get the list of all memory allocations and de-allocations during program execution, 
along with the virtual addresses of the allocations.

\begin{figure}[t]
\vspace{3pt}
\centering
 \includegraphics[width=0.48\textwidth]{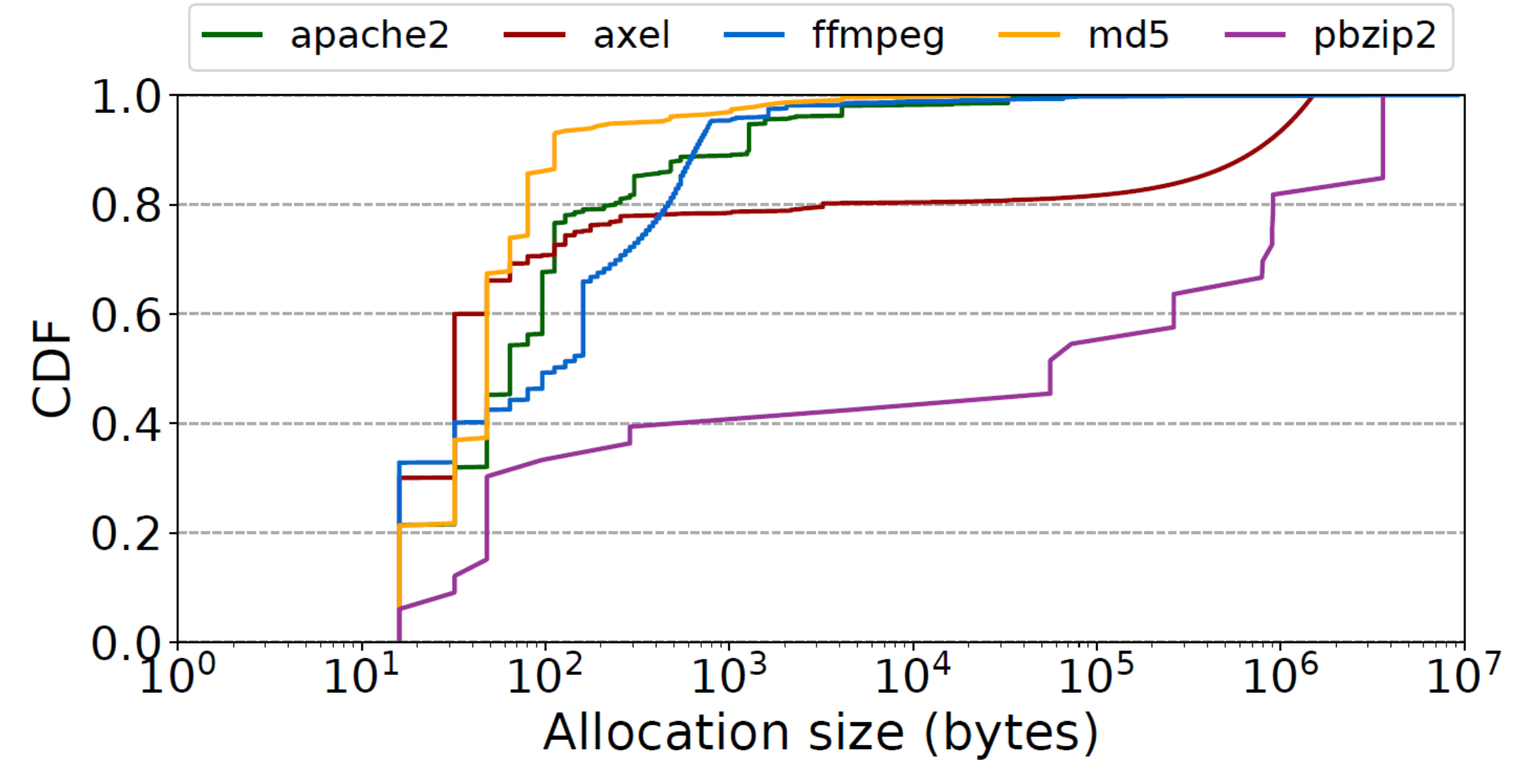}
 \vspace{-17pt}
    \caption{CDF of allocation sizes for all five workloads. The x-axis denotes allocation size in log-scale.}
\label{fig:workload-alloc-cdf}
\end{figure}

Figure~\ref{fig:workload-alloc-cdf} plots the CDF of allocation sizes for all five workloads. 
Table~\ref{tab:workload-stats} lists details of the five workloads, 
along with the number of allocations, average allocation sizes, and the total number of pages used. 
We also report the average number of runs per page, 
calculated by first measuring the maximum number of runs on each page, during the program execution, and then averaging this over all pages.

We remark that to convert the virtual addresses of allocations to physical page addresses we used 4KB pages.
While Linux exposes virtual to physical page mappings via {\tt /proc/PID/pagemap}, 
it can be unreliable due to lazy mapping, copy-on-writes, and page swapping.
We therefore did not use it. 
We discuss how page size impacts our results in Section~\ref{sec:btree-discussion}.



    
    

\begin{table}
\resizebox{0.48\textwidth}{!}{
\begin{tabular}{|l|l|l|l|l|l|}
\hline
Name    & Description                                                                 & \#Alloc. & Avg Size & \#Pages & \begin{tabular}[c]{@{}l@{}}Avg \#Runs \\ Per Page\end{tabular} \\ \hline
apache2 & \begin{tabular}[c]{@{}l@{}}Apache bench \href{https://httpd.apache.org/docs/2.4/programs/ab.html}{ab} \\ 20k req.\end{tabular}             & 771      & 1035                                                      & 207     & 6.46                                                           \\ \hline
axel    & \begin{tabular}[c]{@{}l@{}}Download\\ 6KB file\end{tabular}                 & 231      & 144745.7                                                  & 8918    & 4.87                                                              \\ \hline
ffmpeg  & Resize image                                                                & 4734     & 5597                                                      & 6545    & 2.9                                                            \\ \hline
md5sum  & \begin{tabular}[c]{@{}l@{}}Calculate md5 \\ hash of 0.5GB file\end{tabular} & 231      & 265.3                                                     & 19      & 18.32                                                          \\ \hline
pbzip2  & \begin{tabular}[c]{@{}l@{}}Compress 2.5MB file\\ 1thread\end{tabular}       & 34       & 818859.7                                                  & 6800    & 2.01                                                           \\ \hline
\end{tabular}
}
\vspace{5pt}
\caption{\rm Details and Statistics of 5 real-world workloads collected using valgrind.}
\vspace{-20pt}
\label{tab:workload-stats}
\end{table}



For FFmpeg, out of 4734 allocations, Valgrind was unable to track 21 allocations of total size 1848 bytes. We manually insert these missing allocations by simulating the worst possible workload for us: we add another page, with 20 allocations of size 16 bytes and 1 of size 1528 bytes, with 16 bytes of metadata between each allocation. 





\paragraphb{Setup} 
Before we describe the results, let us briefly explain our setup.
We simulate the 5 workloads on a 2.11GHz x86 machine, based on Glibc and Linux's current implementations of memory tagging. 

As mentioned in \S\ref{sec:btree-impl}, we start with reserving a buffer to store the tags, of the same size as the 4-bit tag array, i.e. 2GB for 64GB RAM. 
We initialize the BTree for each page, with only 1 run starting at 0 and tagged with 0. This mimics Linux's behavior of resetting all memory tags to 0 at initialization. We simulate the workloads by iterating through every malloc/free call as recorded by Valgrind. 

For every new allocation, we call {\sf IRG} (implemented in software using rand()) to generate a new random tag, and then call our {\sf STG} implementation to update the BTrees in the corresponding pages, to tag all the granules in the allocation with the generated tag value. All the granules between allocations are tagged with zero since Glibc tags its metadata with zero. This means that there is a 0-tagged run preceding every allocation with a non-zero tag.

For every de-allocation (free or delete call), we iterate over all the freed granules and call {\sf STG} to update their tags in the corresponding pages' BTrees to zero. This is in line with Glibc's approach of tagging freed memory with zero. We note that our {\sf STG} function tries to merge adjacent runs with the same tag, so freed allocations' runs might be merged with nearby metadata runs since both are tagged with zero.

If at any point during the simulation, the BTree for a page tries to use more space than 128 bytes (reserved buffer space per page), we stop updating the BTree and use an array instead. Lastly, to ensure correctness, we call {\sf LDG} and check the tags of all granules after every allocation/de-allocation call.


\begin{figure}[t]
\centering
 \includegraphics[width=0.48\textwidth]{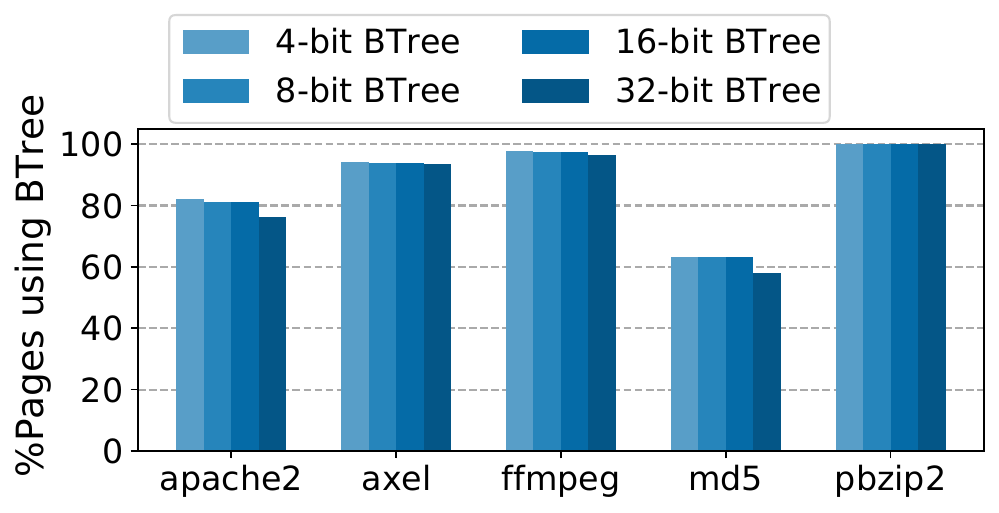}
 \vspace{-17pt}
    \caption{Proportion of pages that use the BTree (and never switch to tag array) for 4-bit, 8-bit, 16-bit, and 32-bit tags, across all five workloads. (Higher is better)}
\label{fig:workloads-pages-using-btree}
\vspace{-10pt}
\end{figure}

\begin{figure}[t]
\centering
 \includegraphics[width=0.48\textwidth]{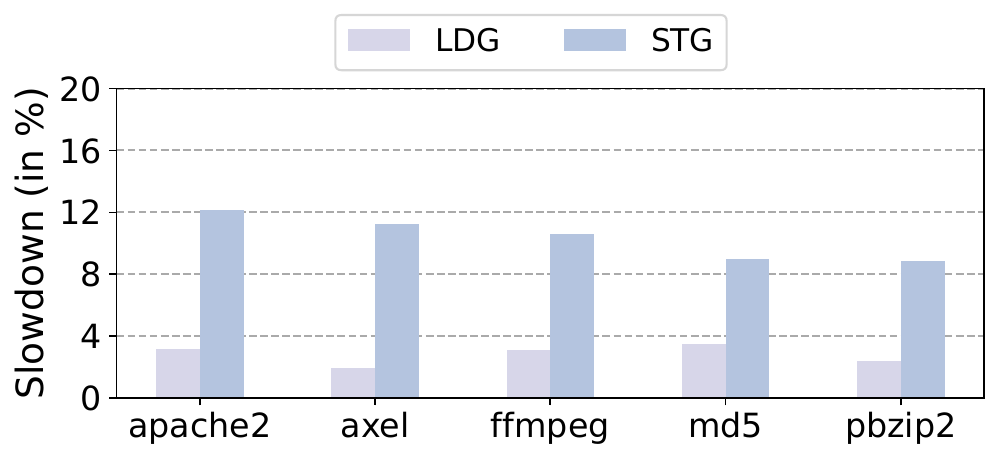}
 \vspace{-17pt}
    \caption{Performance overhead (in \%) of the LDG, STG operations on BTree, relative to tag array, both storing 4-bit tags.
    Smaller is better.
    These numbers are based on the average time for these operations across all pages. Note that this is based on an unoptimized, software implementation, and is not representative of a hardware-based implementation.}
\label{fig:workloads-perf}
\vspace{-10pt}
\end{figure}

\paragraphb{The results} 
We compare our data structure to the existing approach of a 4-bit tag array, with respect to 
(i) supporting larger tags within the same amount of space, and 
(ii) Performance: we report the latency of {\sf STG} and {\sf LDG} in our software implementation, along with tree depth.




\paragraphi{Supporting Larger Tags} 
Figure \ref{fig:workloads-pages-using-btree} plots the
fraction of pages that do not switch to the 4-bit tag array
in our adaptive design with the BTree storing 4-bit, 8-bit, 16-bit, and 32-bit tags, for all workloads.
We make the following key observations from the results:
\begin{itemize}[leftmargin=1em]
    \item The per-page BTree fits within 128 bytes for 57.9\% to 99.98\% pages, across all workloads and all tag sizes.
    This means that
    our design is able to support 8-bit, 16-bit, and 32-bit tags for 57.9\% to 99.98\% of all pages while using the same memory overhead of 3.125\%. 
    Hence, using our approach, we can get stronger security guarantees by using larger tags for at least 57\% (and up to 99\%) of pages, across a wide variety of real workloads.
    
    \item For the Apache, FFmpeg workloads, on varying the tag size from 4-bit to 32-bit, there is a 6\% decrease in the proportion of pages that do not switch to the tag array. This is because, while the structure of each BTree remains the same, the space used by each node in the tree increases with tag size (as described in \S\ref{sec:btree-analysis}). Hence, more pages switch to the 4-bit tag array as their BTrees do not fit within 128 bytes for larger tags. This is expected as per our analysis in \S\ref{sec:btree-analysis}.
    \item Across different workloads, we observe that the proportion of pages that do not switch to 4-bit tag array increases as the average number of runs per page increases. 
    For instance, md5sum has the largest average of 18.3 runs per page, as compared to pbzip2 which has the smallest average of 2 runs per page. Hence, the md5sum workload has only about $60\%$ of pages that use the BTree. This corroborates our analysis (\S\ref{sec:btree-analysis}) since, for a larger number of runs per page, more pages would have to switch to the tag array as their BTree would not fit within 128 bytes.   
\end{itemize}

\paragraphi{Performance} For 4-bit tags, LDG instructions in our software implementation take up to $3\%$ more than a basic tag array on average, 
and STG takes up to $12\%$ more. This is across all workloads. Figure~\ref{fig:workloads-perf} plots the average performance of our 4-bit BTree relative to a 4-bit tag array. Note that these numbers are based on our software implementation and are not representative of an ARM hardware implementation. 
The average depth of the BTree is between $1$ and $1.17$ for all workloads and tag sizes, with the max depth being 3 for 4-bit tags and 2 for 8-bit to 32-bit tags.

\subsection{Discussion}
\label{sec:btree-discussion}
\paragraphb{Effect of Page Size} For our evaluation, we assumed that all pages are small, i.e. of size 4KB, but Linux supports using huge pages of sizes 64KB -- 1GB on arm64. 
With larger page sizes, we would expect similar or better results because (i) large allocations will be split up into lesser pages, leading to lesser runs per page on average, (ii)
as the buffer space for each page will increase proportional to the page size, we will be able to fit the BTrees for pages with larger number of runs as well, e.g. for pages of size 64KB for 4-bit tags, each node will be of size 23 bytes (with 16-bit keys, and child, parent pointers), and the buffer space will be 2KB, meaning that BTrees with over 80 nodes can fit within this space before overflowing. Hence, we will be able to support BTrees with large tags for even more pages than our results for the 4KB page size.


\paragraphb{Prefetching Tags} MTE supports prefetching tags when cache lines are loaded for tagged memory. We can also support efficient pre-fetching of tags, with runtime similar to the {\sf LDG} operation, by also fetching information about neighboring runs when searching for a granule in the tree.

\paragraphb{Page Swapping} When swapping pages in/out of the disk, the 128-byte buffer corresponding to the page being swapped (containing BTree or tag array) can be copied to/from the disk, along with the page.


\paragraphb{Supporting Large Pointer Tags} Our design allows us to use much larger tags with limited memory usage. But, we are still limited by the size of the pointer tag. Currently, ARM supports TBI~\cite{linux-tag-addr-abi}, allowing the top 8 bits of pointers/virtual addresses to be used to store tags or other metadata. Also, Pointer Authentication Code (PAC)~\cite{arm-pac-whitepaper}, which was added to Armv8.3 for control flow integrity, is allowed to use the top 3-31 bits of the pointer, based on the processor's virtual address size.
To be able to use larger pointer tags for MTE, we could use all 8 bits in the top byte for MTE, or extend TBI to use some PAC bits for MTE. We discuss the possibility of combining MTE with PAC in \S\ref{sec:improve-combine-pac-mte}, to allow using up to 31 bits for memory safety.
We could also use the fat pointers idea wherein the tag is stored in another word and passed along with the pointer for address translation~\cite{cheri-isca14, lowfat-ptr}.


\paragraphb{Supporting LDGM/STGM for User space} Currently, bulk tag manipulation ({\sf LDGM/STGM}) instructions are supported only in kernel space. But our BTree design can do bulk tagging almost as efficiently as a single {\sf LDG/STG} call -- by getting/setting tags of a run (instead of a granule). Hence, it might be possible to support this instruction for user space as well.

\paragraphb{An attack: forcing pages to switch to a tag array} 
If an attacker could exploit a vulnerability and use techniques like heap grooming to cause a large number of runs on a single page, 
the system will switch that page to a tag array with short 4-bit tags.
Security for allocations on that page will be reduced to the security of current MTE, namely $1/16$ chance of guessing a tag. 
However, in a system that uses a randomized allocator~\cite{autoslab-heap-spray}, with features such as, (i) allocations are assigned random locations from the free list, (ii) separate blocks of memory for objects of different objects and/or sizes,
it can be difficult for an attacker to cause many runs on a single page.

\section{Additional improvements to MTE}
\label{sec:additional}

\noindent
Finally, we discuss a few additional enhancements to MTE. 

\subsection{H/W Supported Invalid Tags}  
\label{sec:improve-invalid-tag}

In Section~\ref{sec:mte-impl-scudo} we saw that tagging freed memory with a random tag helps detect Use-after-free's with 93\% probability. 
This can be further improved if the hardware supported an explicit {\em invalid} tag:  a tag that causes every access to the granule to raise an exception.
A freed granule can be tagged with this tag so that the hardware will raise an exception on any attempt to access this granule. 
Since the invalid tag value cannot be used to tag allocated memory, this will leave 15 possible tags for allocated memory (assuming 4-bit tags are used).
We note that granules used to store metadata need to be accessible to the allocator and hence will have to be tagged with a random valid tag value.
The invalid tag is akin to an unreadable guard page, however, assigning the invalid tag to a freed granule allows access control at a much finer granularity.

\subsection{Combining PAC and MTE for Memory Safety} 
\label{sec:improve-combine-pac-mte}

Recall that ARM PAC (Pointer Authentication Code) embeds a cryptographic integrity tag (a MAC) into the high-order bits of every pointer and return address. This means that for data pointers, we need to store both the MTE tag and the PAC in the pointer.
A 64-bit pointer has a limited number of unused bits for these tags: 
only 11 to 31 bits can be used to store both the PAC and MTE tags
(32 to 52 low-order bits hold a virtual address and 
the 55th bit is reserved for selecting the high or low half of the virtual address space).

Concretely, for a machine with 47-bit virtual address space, there are \combinedtagsz unused bits; we could either use all \combinedtagsz bits for MTE tags (using our design) or use all the bits for PAC.
More generally, we can use $x$ bits for MTE and $\combinedtagsz-x$ bits for PAC for some $0 \le x \le \combinedtagsz$. 
This presents a tradeoff between memory safety and pointer integrity for data pointers. 
 
 We propose to improve this significantly by combining MTE and PAC for data pointers. 
 By storing a single \combinedtagsz-bit tag in every data pointer we show 
 that the combination provides memory safety similar to a \combinedtagsz-bit MTE 
 and pointer integrity similar to a \combinedtagsz-bit PAC for most of RAM.

\paragraphb{Design} 
Let $k$ be a PAC secret key known to the processor.
Whenever a new allocation is made at address $a$,
the memory allocator generates a fresh random \combinedtagsz-bit MTE tag $t_m$. 
Next, the processor generates a \combinedtagsz-bit PAC tag $t_p$ as follows:
\[
    t_p \gets \text{QARMA}\bigl( k, (t_m, a) \bigr).
\]
Here QARMA is the block cipher used in PAC (128-bit key and 64-bit block size),
where the output of QARMA is truncated to \combinedtagsz bits.
This $t_p$ is embedded in the high-order bits of the returned pointer.
The system can include additional context (such as the object type~\cite{ccfi,pac-parts-sec19,pac-security22}) 
as further input to the cipher.

In addition, for every allocation, the processor stores the allocation's starting address $a$ and the \combinedtagsz-bit memory tag $t_m$
in the adaptive data structure from Section~\ref{sec:btree-design}.
Hence, this scheme has exactly the same memory overhead as if MTE used 4-bit tags. 
Looking up any address in the data structure returns the starting point $a$ of the enclosing allocation along with $t_m$. 
For pages where 
the system switches to a tag array, the processor stores the lower four bits of $t_p$, which we denote by $t_m'$. 
This is closely related to PACSan's~\cite{pacsan-arxiv} pointer tags, 
but our data structure for storing metadata in memory is different and provides better support for caching, pre-fetching, and page swapping.

At every load/store instruction, the system compares the \combinedtagsz-bit pointer tag $t_p$ to $\text{QARMA}\bigl(k, (t_m, a)\bigr)$ 
where $(t_m, a)$ are obtained from the page's BTree.
Here $a$ is the starting address of the allocation containing the address in the pointer,
and $t_m$ is the MTE tag of that allocation.
If the page has been switched to a tag array, the system compares the lower four bits of $t_p$ to the 4-bit $t_m'$ stored in the tag array. 
In this case, the scheme reduces to regular MTE.

\ignore{
Note that we are comparing 16-bit tags, even though only a 4-bit $t_m$ is stored in memory 
(the address $a$ is always stored in the BTree irrespective of the tag length). 
}


\begin{table}
\begin{center}
\vspace{7pt}
\resizebox{0.48\textwidth}{!}{
\begin{tabular}{l|l|l|l}
         & \begin{tabular}[c]{@{}l@{}}MTE \combinedtagsz-bit\\ $|t_m| = \combinedtagsz$\end{tabular} & \begin{tabular}[c]{@{}l@{}}PAC+MTE\\ $|t_p| = \combinedtagsz = |t_m|$\end{tabular} & \begin{tabular}[c]{@{}l@{}}PAC \combinedtagsz-bit\\ $|t_p| = \combinedtagsz$\end{tabular} \\ \hline
Memory Overhead     & (\combinedtagsz,$|a|$)  & (\combinedtagsz,$|a|$)     & 0   \\ \hline
Linear Overflow     & 99.9   & 99.9  & 0    \\ \hline
Non-Linear Overflow & 99.9   & 99.9   & 0    \\ \hline
Use-after-free      & 99.9   & 99.9    & 0     \\ \hline
Use-after-realloc   & 99.9   & 99.9    & 0     \\ \hline
Pointer Corruption  & 0    & 99.9  & 99.9  
\end{tabular}
}
\end{center}
\vspace{10pt}
\caption{\rm Security Analysis of MTE vs. PAC+MTE vs. PAC all using \combinedtagsz-bit tags. We list the memory overhead for storing a single run in the BTree. For every attack, we list the probability (in \%) of a benign or malicious attempt being detected. We assume random re-tagging on free. PAC+MTE provides the benefits of a 16-bit PAC and a 16-bit MTE, while only using 16 bits.}
\label{tab:combine-pac-mte-security}
\vspace{-15pt}
\end{table}

\paragraphb{Security}
Table \ref{tab:combine-pac-mte-security} compares the protection provided by MTE, PAC, and PAC+MTE
against memory unsafety and pointer corruption.
In all three examples we use \combinedtagsz-bit pointer tags.

For overflows into adjacent or non-adjacent allocations, 
MTE can detect both benign and malicious attempts with a probability of at most $65535/65536$.
For PAC+MTE, the probability of the overflow succeeding is $1/65536$.
This is because the attacker can, at best, guess the \combinedtagsz-bit MAC $t_p$ with probability $1/65536$, 
even if they know the starting address ($a$) of an allocation.
Hence, PAC+MTE is able to provide the same spatial safety as MTE.

For Use-after-free, if freed memory is tagged with a random tag $t_m$, 
MTE provides $65535/65536$ protection.
For PAC+MTE, if a dangling pointer (with tag $t_p$) is used to access freed memory, 
the BTree will return the same starting address (since freed memory also forms a run with the same address but just a different tag), 
but the memory tag will be different. 
As a result, a benign or malicious UAF attempt will succeed with $1/65536$ probability. The analysis is similar for Use-after-realloc. PAC+MTE therefore provides the same temporal safety as MTE.




In terms of pointer corruption, an attacker can only guess the correct PAC for a new address with probability $1/65536$. Hence, PAC+MTE provides the same pointer integrity as PAC with \combinedtagsz-bit pointer tags.

In summary, the PAC+MTE scheme uses larger pointer tags, which, combined with the data structure from Section~\ref{sec:btree-design}, 
can provide stronger memory safety and pointer integrity for 57\% to 99\% RAM, with the same memory overhead as MTE with 4-bit tags. Note that pointer reuse attacks are possible in both PAC and PAC+MTE schemes, but they can be made harder by including additional context in the MAC,
such as the allocation size and object type.
\ignore{
Note that for a given address~$a$, the \combinedtagsz-bit $t_p$ is always one of 65536 possible values determined by the \combinedtagsz-bit $t_m$.
If the attacker could obtain all 655 possible values for $t_p$ at $a$, 
then all protection probabilities revert to $15/16$. 
}

\section{Additional related work}
\label{sec:related}

In this section we review more of the related work relating to address sanitization and memory tagging. 

Address Sanitizer (ASan)~\cite{asan-atc-12} is a software-only tool based on compiler instrumentation introduced to the LLVM and GCC compilers in 2011. It uses red-zones around every object for spatial safety, uses quarantines to delay the reuse of freed objects to detect temporal bugs, and also stores 1-byte metadata per 8 bytes of memory to mark memory as valid/invalid. While ASan is heavily used in testing, it cannot be used in production code due to its large CPU and RAM overhead.
Califorms~\cite{califorms-micro19} is a research proposal based on the idea of blacklisting freed memory and memory around allocations, similar to ASan's quarantine and red-zones. This has a high overhead and also cannot protect against non-linear overflow beyond the blacklisted memory. 
MarkUs~\cite{markus-sp20} uses an approach similar to quarantines, but provides no safety against spatial attacks. 


There is considerable prior work on storing metadata to enforce memory safety. 
Fat pointer schemes that maintain such metadata in a separate memory region~\cite{hardbound-asplos08, softbound-pldi09, intel-mpx} incur a large CPU and memory overhead. 
CHERI~\cite{cheri-isca14} is a popular proposal that stores metadata inline with the pointer.  However, it changes the memory layout and loses pointer-size compatibility with legacy code.

In-FAT pointers~\cite{infat-asplos21} provide both inter and intra-object spatial safety using per pointer metadata, but cannot protect against temporal attacks. FRAMER~\cite{framer-acsac19} uses 16-bit pointer tags to lookup object metadata stored near each object, but has large performance overheads due to its software-based implementation. Low-FAT pointers~\cite{lowfat-ptr, lowfat-ptrs-stack} use binned allocators and store the size of the object within the pointer to provide spatial safety, but they have large memory overheads with no temporal safety. CUP~\cite{cup-asia-ccs-18} is another approach that uses the entire pointer to store tags to index into a metadata table storing base and bounds. However, this only provides partial temporal safety and also has a large performance overhead, since the pointer first needs to be translated into an actual address before it can be de-referenced.

AOS~\cite{aos-micro-20} stores a pointer authentication code (based on PAC~\cite{arm-pac-whitepaper}) in the pointer's top bits, and uses it to index into a hash table storing object bounds. It provides partial temporal safety and does not apply if external (unmodified) modules modify a pointer. 
PACSan~\cite{pacsan-arxiv} uses a similar idea, it generates a pseudo-random birthmark when an object is allocated (similar to MTE tags) and uses a PAC code to index into a hash table that stores the birthmark and object bounds. But it is susceptible to use-after-realloc (UAR) because the birthmark is not fully random. It also does not support tag caching, pre-fetching, and page swapping.

There is considerable work on memory tagging. \href{https://clang.llvm.org/docs/HardwareAssistedAddressSanitizerDesign.html}{HWAsan} is a compiler-based tool, that uses 8-bit tags for every 16 bytes of memory. But since it adds extra instructions to check tags before every memory access, it has a $2\times$ performance overhead. The Generic and Software-Tag modes of \href{https://www.kernel.org/doc/html/latest/dev-tools/kasan.html}{KASAN} also use compile-time instrumentation to enforce memory safety using shadow memory, quarantine, and memory tagging.  
UAFSan~\cite{uafsan-issta21} generates a tag at each allocation/de-allocation and stores the tag as metadata for each object and pointer. 
It has considerable memory and performance overhead, and only handles temporal safety. 
All these compiler-instrumentation based tools require re-compilation of source code.

No-FAT pointers~\cite{nofat-isca-21} combine memory tagging with Low-FAT pointers to get temporal and spatial safety. But, 
both No-FAT and Low-FAT require pointers to be within bounds when passed to functions, which is too strict for real-world programs.
CrypTAG~\cite{cryptag-arxiv} combines tagging and encryption -- it stores the tag in the pointer, but, instead of storing the tag for every granule in memory, it encrypts data with the tag as additional input for encryption. 

\cite{msrc-mte-secu-analy} analyses the security of MTE, and recommends that adjacent allocations always have different tags. \cite{msrc-bluehat-mte-attack} discusses a potential attack on MTE, but it is based on the assumption that there is no re-tagging on free; 
this is not the case in the applications we surveyed in Section~\ref{sec:usage-survey}. 
~\cite{tag-guide} presents a survey of tagged architectures and challenges in deploying MTE in the real world. 
While all these works point out issues with MTE, they do not explore the security of software systems using tagging (Section~\ref{sec:usage-survey}) or implementation improvements to MTE,
as in Sections~\ref{sec:improvements} and~\ref{sec:additional}.

HAKC~\cite{hakc-ndss-20} also uses both PAC and MTE to achieve kernel compartmentalization, which is orthogonal to our work. We instead combined them to achieve both memory safety and pointer integrity while using a single 16-bit tag.

MTE does not protect against type errors, and as such, the work on type checking~\cite{effective-san, typesan-ccs16} is complementary to and can be used in conjunction with MTE. 

\section*{Data Availability}
The workload data and source code will be made publicly available via github.


\bibliographystyle{plain}
\bibliography{refs}

\appendices
\section{LLVM Tagging Survey}
\label{sec:app-llvm}
When LLVM creates a new stack frame, it uses the {\sf IRG} instruction to generate a random tag for the base pointer, and then
increments the tag by one for each argument and local variable. 
All stack objects are aligned to 16 bytes (granule size) and the tag value 0 is excluded. 

Note that LLVM does not assign a fresh random tag to each stack variable 
because that would require allocating one register to hold a tagged pointer for each stack variable.
The increment-by-one strategy makes it possible to use the tagged base pointer to access all stack variables using 
pointer arithmetic of the address and the tag, namely using the {\sf ADDG/SUBG} instructions. 

LLVM has not yet implemented tagging for global variables, although there is ongoing discussion on this topic~\cite{llvm-dev-globals-tagging}.
We also note that the granule holding the return address is not tagged because 
it is already protected by ARM's Pointer Authentication~\cite{arm-pac-whitepaper}.

\paragraphb{Security} 
As in the case of the Chrome allocator discussed in the previous section, the deterministic nature of the increment-by-one strategy enables
an overflow attack on stack variables that succeeds with probability one. We were able to successfully run this attack in an MTE-enabled VM in QEMU.

\paragraphb{An improvement}
Our recommendation for using a random odd offset $\delta$ discussed in the previous section 
applies here as well. A different $\delta$ could be sampled for every new stack frame. This can make overflows harder to mount, with much lower register pressure than random tagging for all variables.

\section{Implementation Details}
\label{sec:app-impl}
\paragraphb{Additional Metadata} For each BTree, we also store additional metadata: the root pointer, as a 1-byte value, denoting the relative offset of the root node from the start of the buffer for this BTree. To manage the allocation and de-allocation of nodes as the tree grows, we use a 2-byte bit array to denote which nodes in the 128-byte buffer are free. 16 bits are enough since only 4-9 nodes can fit within the 128 bytes buffer. Lastly, we have an additional 1-bit flag for each page, denoting whether we are using a BTree or an array for storing tags for that page.

\section{Detailed Analysis}
\label{sec:app-analysis}
The total space used by each BTree node is $11 + 4t$ bytes, counting 4 bytes for keys, 5 bytes for child pointer offsets, 1 byte for the parent, 1 byte for the key count and $4t$ bytes for the tags corresponding to 4 keys. Adding the space usage of $m$ nodes in the tree, the root offset, the 2-byte freed nodes bit array and the 1-bit page flag, the total space used by the BTree is given by: 
\begin{equation}
\label{eq:btree-space-usage-1}
    S^{BTree}_t = m*(11 + 4t) + 1 + 2 + 0.125 \text{ bytes}
\end{equation}

We observe that in a 4,5-BTree, (i) each of these $m$ nodes stores at least 2 keys (except for the root which is allowed to have 1 key) and at max 4 keys, and (ii) there is exactly one key per run. Hence, we get $2(m-1) + 1 \leq n \leq 4m$.
Re-writing this to get an inequality on $m$, we get:

\begin{equation}
\label{eq:btree-node-run-eqn}
    \frac{n}{4} \leq m \leq \frac{n+1}{2}
\end{equation}

We can combine Eqs \ref{eq:btree-space-usage-1} and \ref{eq:btree-node-run-eqn} to get bounds on the total space usage of the tree, in terms of $n$ (Eqn \ref{eq:btree-space-bounds}).

For a BTree to fit within this space, the BTree space usage must be less than 128 bytes, i.e. from Eq \ref{eq:btree-space-usage-1}, 

\begin{equation}
m*(11+4t) + 3.125 \leq 128
\end{equation}

This gives us the bounds on $m$ and $n$ as seen in \S\ref{sec:btree-analysis}.

\section{Evaluation Details}
\subsection{Workloads}
We used Valgrind to get the list of all memory allocations and de-allocations during program execution. An example command for profiling the apache2 program is ``valgrind -{}-trace-malloc=yes -{}-trace-children=yes -{}-log-file=log.out /usr/sbin/apache2 -X''.

\paragraphi{Apache} We ran the HTTP-based apache server using ``/usr/sbin/apache2 -X''  and used the Apache benchmark library \href{https://httpd.apache.org/docs/2.4/programs/ab.html}{ab} to issue 20,000 requests to the server (``ab -n 20000 http://127.0.0.1/'').

\paragraphi{Axel} \href{https://linux.die.net/man/1/axel#}{Axel} is a light-weight download accelerator for Linux. We ran ``axel https://github.com/llvm/llvm-project/blob/main/README.md''. 

\paragraphi{Ffmpeg} We ran the command ``ffmpeg -i IMG\_8554.jpg -vf scale=360:240 out.png'' to profile ffmpeg.

\paragraphi{Md5sum} We used md5sum to calculate the hash of the Anaconda installation script, i.e.. ``md5sum Anaconda3-2021.11-Linux-x86\_64.sh''.

\paragraphi{Pbzip2} We use pbzip2 to compress a file, via the command ``pbzip2 -p1 IMG\_8554.jpg''. We run it in single threaded mode since valgrind is unable to track allocations for multi-threaded programs.

\subsection{Memory Usage Results}

\begin{figure}[t]
\centering
 \includegraphics[width=0.48\textwidth]{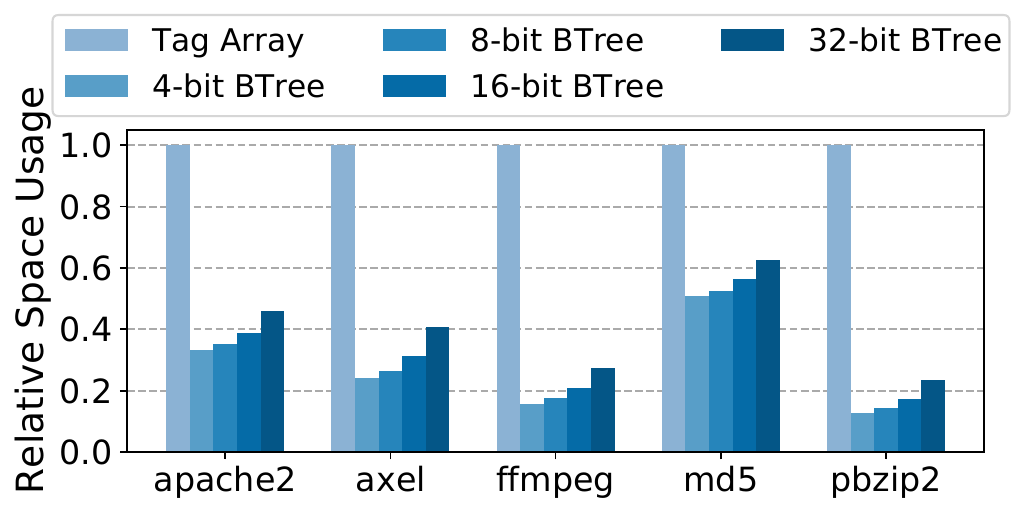}
 \vspace{-17pt}
    \caption{Total Space Usage of our dynamic design, relative to 4-bit tag array. We add the size of the BTree for each page, to get the total space usage, and take the maximum of this value over the course of program execution. (Lower is better)}
\label{fig:workloads-btree-space-relative}
\vspace{-10pt}
\end{figure}
We also measure the total space usage of our design for all pages, by adding the size of all pages' BTrees (or tag arrays) at the end of the workload simulation.
Figure \ref{fig:workloads-btree-space-relative} plots the total space usage for all pages, relative to a 4-bit tag array. Our design uses $0.1\times$ to $0.61\times$ lesser space as compared to the 4-bit tag array, across all workloads as well as tag sizes. Hence, an alternate implementation of the BTree design wherein we do not reserve 128 bytes beforehand for each BTree, can save memory overhead by upto $10\times$ while supporting 4-16 bit tags for atleast 60\% pages. This might involve non-trivial memory management for the B-Trees as the number of runs per page varies, hence we leave further analysis to future work.

\section{Discussion}
\paragraphb{Random tagging for metadata and at free} One of our suggested improvements is to always tag freed memory and metadata with a new random tag. Observe that in our current results, each allocation and each 0-tagged metadata between allocations, forms a separate run, with metadata runs potentially merged with 0-tagged freed granules. 
If freed memory and metadata were tagged with random tags instead, we will still have a similar number of runs on each page, and metadata runs might be merged with nearby granules if tags match, but we won't be able to merge freed granules with metadata. 
Hence, we expect our results to be mostly similar for this implementation.

We note that, if metadata between objects is tagged randomly, it is possible that metadata might end up having the same tag as the adjacent object when the page switches to 4-bit tag array. We observe that this is a possibility if metadata was tagged randomly even in the current MTE design. To avoid this, the allocator could choose a random tag for metadata such that the lower 4 bits are different from those of adjacent allocations.

\paragraphb{Allowing applications to opt in for BTree}
Since we use a dynamic data structure on a per page basis, we can allow applications to opt in for using BTrees for their pages, if potential performance overheads are acceptable for stronger security.

\paragraphb{Tagging in CMAs} Certain applications such as apache and nginx use custom memory allocators (CMA) on top of Glibc. For example, apache uses Glibc to allocate large pools of memory, which are then managed by a pool-based \href{http://www.apachetutor.org/dev/pools}{CMA}. In our evaluation, we assumed that only Glibc handles memory tagging. 
But, supporting tagging in such CMAs can provide memory safety for the smaller allocations within memory pools. We did not explore this in our evaluation since this is not currently implemented.

\end{document}